**Barriers and Solutions to the Adoption of Clinical Tools for Computational Psychiatry**

David Benrimoh[1], Victoria Fisher[2], Catalina Mourgues[2], Andrew D. Sheldon[2], Ryan Smith[3], Albert R. Powers[2*]

1. McGill University School of Medicine, Montreal, Canada

2. Yale University School of Medicine and the Connecticut Mental Health Center, New Haven, CT, USA

3. Laureate Institute for Brain Research, Tulsa, OK, USA

*Correspondence should be addressed to:

Albert R. Powers, M.D., Ph.D.
The Connecticut Mental Health Center, Rm. S109
34 Park Street
New Haven, CT 06519
albert.powers@yale.edu
203.974.7329

**Abstract**

Computational psychiatry is a field aimed at developing formal models of information processing in the human brain, and how alterations in this processing can lead to clinical phenomena. Despite significant progress in the development of tasks and how to model them, computational psychiatry methodologies have yet to be incorporated into large-scale research projects or into clinical practice. In this viewpoint, we explore some of the barriers to incorporation of computational psychiatry tasks and models into wider mainstream research directions. These barriers include the time required for participants to complete tasks, test-retest reliability, limited ecological validity, as well as practical concerns, such as lack of computational expertise and the expense and large sample sizes traditionally required to validate tasks and models. We then discuss solutions, such as the redesigning of tasks with a view toward feasibility, and the integration of tasks into more ecologically valid and standardized game platforms that can be more easily disseminated. Finally, we provide an example of how one task, the conditioned hallucinations task, might be translated into such a game. It is our hope that interest in the creation of more accessible and feasible computational tasks will help computational methods make more positive impacts on research as well as clinical practice.

## Introduction

A major area of both need and opportunity in the field of psychiatry is establishing the link between observed symptoms (e.g., the criteria we use to diagnose and characterize illnesses) and neurobiological findings (e.g., alterations in functional connectivity or gene expression) via alterations in established information processing carried out by the brain. The challenge posed here is that of mapping symptoms onto neurobiology in a principled manner, based on a sound understanding of what the brain is computing, how this computation is implemented neurobiologically, what specific computations are altered in disease, what changes in neurobiological processes account for these alterations, and how these alterations give rise to symptoms. This work is carried out in the hope that an improved mechanistic understanding of psychiatric illnesses will lead to new treatments and biomarkers that could be linked to treatment mechanisms of action.

The field of computational psychiatry aims to fill this need [1–4]. There are many proposed computational approaches for studying relevant psychiatric problems, ranging from reinforcement learning models and hierarchical bayesian models to data-driven methods, such as deep learning. However, common to each is the aim of identifying latent states that drive both normative brain function as well as symptom and disease development: as in physics, we should be able to formally describe models by which the brain processes information and then design experiments and gather data that specifically support, refute, or call for a modification of the models proposed. The key element of computational psychiatry is that many of these models, which can be used to simulate behavior and which can be fit to observed data, contain parameters corresponding to latent states or processes that are not otherwise easily observable. These parameters can then in turn be correlated not only with behavior, but with various kinds of neural measures [5-7], ultimately linking behavior and neural implementation via computation. This approach has been applied to a number of conditions such as psychosis [5, 8, 9, 10,11,12,13], anxiety and depression [7,14,15,16], obsessive-compulsive disorder [17], substance use [18, 19], and transdiagnostic samples [20, 21, 22] and remains an area of emphasis for major funding sources in psychiatric research, including the National Institute for Mental Health (NIMH). Indeed, computational approaches are viewed by some as the field's best option for bringing psychiatric nosology into step with that of the rest of medicine, by linking disease manifestations and distal etiologies via distinct mechanisms [2,3].

It is worthwhile discussing one illustrative example of this approach that has recently been applied to understanding hallucinations, which may be formulated as arising because of a tendency to over-estimate the reliability of one's prior expectations (or *priors*, in Bayesian terms) during perception [5,9,11,12]. With this hypothesis in mind, researchers formulated a task to create conditioned hallucinations (CH) by heightening expectations [5]. CH occur as a result of classical conditioning, where a subject is presented with a salient stimulus paired with a difficult-to-detect target (e.g., an image and a sound) at the same time in a repeated manner, such that in the presence of the salient stimulus (e.g., the image) and the absence of the target, the subject may hallucinate the target due to their strong expectation that the stimulus should be present. This task is modeled using the Hierarchical Gaussian Filter or HGF [23] which estimates parameters that can in turn be associated with neurobiological measures, such as imaging [5]. We will return to this illustrative example later in this article.

So why have computational methods not been adopted into current large-scale research efforts or clinical practice? Below, we will briefly discuss some of the barriers to widespread adoption that computational psychiatry faces, as well as some potential routes to overcoming them.

*Barriers*

We propose that practical concerns are most likely to act as barriers to the widespread implementation of computational approaches in psychiatry. Most important among these are time, ecological validity, and practical implementation concerns.

*Time:* When designing or modifying a task intended to be suitable for computational modeling, the primary concern of the designer is generally the validity of the task in terms of its ability to capture relevant latent states that drive behavior. This is similar to the problem faced by many classical neuropsychological or psychophysical tests, which often require many trials to establish reliable measures and which in turn can require, in some cases, one to multiple hours of testing, depending on the range of tasks included [24, 25, 26]. Recently-published task-model combinations in computational psychiatry take between 15 and 40 minutes for each assessment [10, 27, 28, 29, 30]. Adding any *one* of these tasks may present a burden in the context of a larger study that must also collect various clinical and neurophysiological measures. In addition, each of these tasks is optimized for a certain set of parameters; as such, multiple tasks would likely be required for the recovery of an adequate computational phenotype of an individual. As these tasks have not yet been integrated into batteries optimized for feasibility, larger-scale research projects would be hard-pressed to include several of them into their protocols.

*Ecological Validity:* Another significant limitation of current tasks in computational psychiatry, as well as in more traditional neuropsychological testing, is the fact that they are not ecologically valid: behavior or experiences during a task most often do not reflect clinically relevant content domains, contexts, and/or symptoms as they are experienced in the real world [31]. For example, decisions aimed at maximizing small amounts of monetary reward or minimizing small shocks in the lab could plausibly engage very different prior expectations than decisions in real-world contexts to avoid feared situations or to maximize overall life satisfaction. Laboratory tasks are also generally designed to isolate certain behaviors or cognitive skills so that they can be effectively measured, modeled, and interpreted. Unfortunately, this ignores the fact that, during real-world functioning, a participant might employ multiple skills when solving a given problem or might need to solve different problems in sequence or in parallel; additionally, various affective or memory-based cues, or volatility in the environment, may interfere with function in a manner not apparent in the more sterile environment of a task [31, 32, 33]. While some neuropsychological and computational tasks (and their parameters) have been linked directly to patient experiences and outcomes, such associations have been limited to date [3,18, 19, 20, 21, 22].

*Implementation Concerns:* In addition to the fact that a standard battery does not yet exist that would facilitate the adoption of these measures, there are several other practical barriers to implementation. One is the fact that computational psychiatry remains a niche field, with few investigators being equipped to implement, refine, and interpret relevant models. This is problematic because, in many cases, investigators with relevant research questions, but without a computational background, would benefit from being able to implement a standardized version of a task that produces an output with a clear report of the results, but are prevented from doing so because user-friendly versions of tasks or models often do not exist. In the neuropsychological realm, this is a problem that is partially addressed by digitized batteries, where investigators who may not be experts in the development or implementation of cognitive tests can still make use of a standardized testing platform and utilize the results [25]. For those investigators who do have a computational background, or an interest in developing relevant expertise, the process of developing and iteratively validating models and tasks can also be prohibitive in terms of both time and expense, resulting from the large sample sizes required for testing.

*Test-retest reliability:* To be useful as clinical assessment tools, it is also imperative that computational model parameters can be measured repeatedly over time (e.g., at key points during treatment). Yet, as many of these tasks involve learning, decision strategies can change (e.g., become more habitual and with more confident priors) with repeated performance. This can result in poor test-retest reliability, raising the concern that the same latent computational process is not being captured at each timepoint of assessment.

*Solutions*

We argue that each of the barriers above have arisen because the field of computational psychiatry has not yet pivoted from validation of constructs to implementation of tools. We propose the following solutions to address these barriers in turn.

*Time:* In order to reduce the time required to complete a given task, existing tasks should be redesigned with a focus on determination of the most efficient structure possible, to allow for reliable parameter estimation in the shortest amount of time. For example, it was recently determined that separation between hallucinators and controls occurs fairly early on in the conditioned hallucinations task, information which is now being used to generate a shortened version of this task [27]. The CH task's relatively simple structure (i.e., establishing a prior and then gradually testing its strength) made a simple empirical review of the data sufficient to determine when the task could be reasonably truncated. However, other task designs may differ in ways that render this determination more complex. In these situations, simulations of data generated by a given task could be performed with the specific aim of determining how many trials are required to derive parameters of interest with tolerable accuracy; indeed, simulations aimed at generating data which can then be compared to the data produced by participants is regarded as being an important part of good practice and model validation in the field of computational psychiatry[3]. As these tasks are shortened, it will also be important to consider the tolerability of these tasks in aggregate, if a traditional battery structure is to be considered. One way to improve tolerability and increase engagement would be to incorporate these tasks into the structure of a game, a strategy used in other disciplines, such as education [34]. This idea of a computational battery constructed in the form of a game is one that we will continue to develop.

Reducing the time required to complete each task would be a useful first step. Another opportunity comes in designing novel tasks meant to model behavior driven by several latent states, where several tasks are constructed in order to have some redundancy in the latent states estimated. This would potentially allow for fewer trials of each task but provide enough information to derive the latent states common to the tasks. These tasks could then be tested in simulations to determine the optimal number of trials thought to be needed, and this could then in turn be tested empirically to determine if the number of shortened trials produces valid estimates compared to longer versions of the tasks. An example of shared parameter estimation would be of two tasks, one focused on perceptual judgements and the other on social judgements. While these two processes likely engage some independent processes, there are likely to be some common parameters shared between them, especially if symptoms exist that seem to affect both domains (e.g. paranoid hallucinations about a neighbor co-occuring in someone with paranoid beliefs about their neighbors). In estimating parameters using information from both tasks, it may be possible to efficiently estimate the shared parameters–and at the same time to determine which parameters are *not* shared, which in and of itself would be an interesting mechanistic finding.

The use of explicit computational models is actually helpful in this case: generative models of behavior are explicitly designed to account for and measure different latent states driving behavior that can appear similar when analyzed via descriptive summary statistics. Thus, in principle, the use of explicit models capable of taking into account multiple drivers of behavior would make for a maximally efficient route toward estimating as many latent states as possible at any given time.

*Ecological Validity:* The idea of integrating tasks within a game that has a believable world, perhaps one modeled on the experiences of either the general public or specific groups of participants, may also have benefits with respect to ecological validity. By integrating tasks into this "gameworld", it would allow for participants to use multiple skills or be influenced by previous experiences within the game, while solving problems reminiscent of those they have to solve in clinically relevant real world contexts (e.g., probing issues of avoidance, approach-avoidance conflict, proximal vs. distal planning, exploration vs. exploitation with respect to social contingencies and their volatility, etc.). This in turn may increase the generalizability of results from these tasks to clinical outcomes of interest. There is also an opportunity here that goes beyond simply improving the ecological validity of current tasks: by creating a gameworld, participants would be able to make choices, approach problems in different ways, and generally act with greater agency than is possible in isolated computational tasks. This in turn would allow for the modeling of new parameters related to patient choice and their generation of action plans; these parameters in turn may reflect latent states more relevant for the generation and maintenance of various symptoms and syndromes [35], but which have not been previously measured in ecologically valid ways.

*Implementation Concerns:* The generation of an integrated battery of efficient computational tasks may also help address some of the concerns around implementation. Firstly, a more comprehensive gameworld (with multiple nested perceptual and decision tasks) would provide a standardized environment, and could be engineered to elicit reports of participant behavior based on computational models programmed into the software. The nested tasks and models could also be made modular and modifiable, to support researchers with various levels of computational experience in their use of this methodology. Digital tools, such as games, are explicitly designed to be easy to disseminate and require minimal training for participants, with training often able to be delivered in the format of a tutorial experience within the game. Hosting these tools online would allow access to large populations of participants who otherwise would not be reached by current lab-based efforts. As such, the expense required for the development and validation of computational tasks and models would be significantly reduced, and their use would be feasible for a larger number of researchers with varying levels of expertise and resources.

*Test-retest reliability:* One solution is that tasks be vetted for test-retest reliability prior to inclusion in battery. In previous research, some have been shown to exhibit higher reliability than others, and depending on the time elapsed between assessments [18, 22]. Another possible solution would be to have participants complete task multiple times before starting to use them for assessment, which could increase reliability if it allows participants to first settle into a stable strategy–which could better mimic the stable strategies they settle into when solving real-world contexts of clinical relevance. Yet another strategy to consider could be to continue to vary task contents, while keeping the abstract decision structure identical. This could in principle minimize changes in initial strategy if participants understood each of these tasks to be new.

*Conclusions and Future Directions*

In this article, we have examined the time requirements, lack of ecological validity, test-retest reliability, and practical considerations such as the lack of widespread computational expertise and the need for expensive validation procedures have limited the utilization of computational psychiatry methodologies in large research initiatives. These barriers must be overcome in order for the utility of computational psychiatry–that is, an improved understanding of latent states and the derivation of parameters that can be linked to neurobiological measures in order to improve mechanistic understanding– to be realized. Our proposed solutions, which are by no means comprehensive or definitive, have focused on a combination of a focus on good design (i.e., re-examining tasks and shortening them when possible, vetting or adjusting tasks to ensure test-retest reliability) and an exploration of the opportunity presented by the integration of tasks into an overarching game world with nested perceptual and decision problems. Depending on its specified contents, such a gameworld could allow for greater ecological validity and the measurement of novel computational parameters, while also providing a modifiable platform that would empower researchers with varying levels of expertise and resources to begin to engage with computational psychiatry. Future work would therefore naturally be focused on the development and validation of this type of broader game environment. While this effort is in its early stages, we present here an example of how a well-validated computational task might be transported into a game world. In **Figure 1**, we see both a depiction of a possible gameworld and a version of the conditioned hallucinations task, where a player learns an association between an auditory and visual stimulus (in this case, a dragon and its growl) while navigating the gameworld, and must react in a way that may affect their gameplay (for example, if they fail to dodge when the dragon is present and growling, they may suffer a penalty). This task is modeled using the HGF, (**Fig. 1B**), in a manner consistent with the gameworld. Here the player is learning the task while playing the game, and their behavior is guided by the logic of the gameworld, rather than it being dictated by the instructions of an arbitrary task. It should be noted that this example is intended for illustration and could be altered if greater relevance to everyday life is required (e.g., replacing the dragon with a more realistic threat).

It is our hope that this article, and the example above, sparks greater widespread motivation towards developing more accessible computational psychiatry measures, bringing them into the mainstream of psychiatric research where they are most likely to have a positive impact on patient care and preventative efforts. Indeed, beyond their use as research tools, it is our hope that something like the gameworld we have briefly illustrated here could eventually become commonplace in the assessment of, and screening for, psychiatric conditions. Furthermore, as has been demonstrated by recent approvals of digital therapeutics for psychiatric indications, these games may also have the potential to serve as accessible and personalizable vehicles for the delivery of treatments.

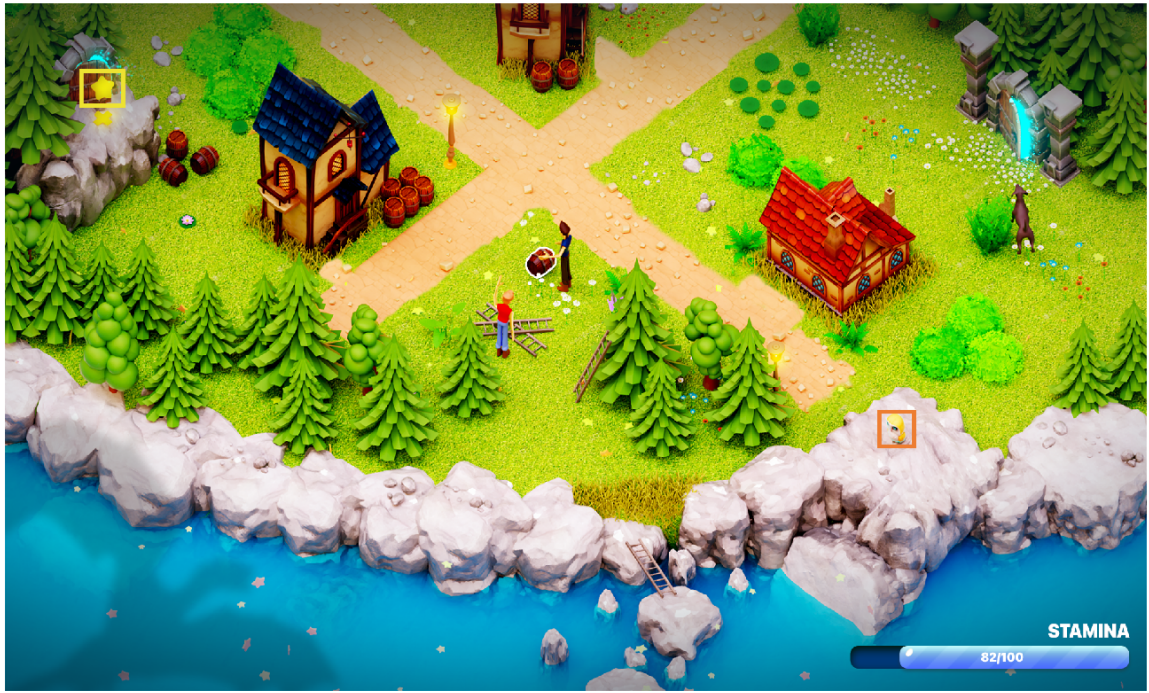

a

b

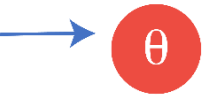

Level 3
Volatility at Level 2

Growl 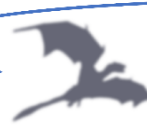

θ

Volatility Evolution Rate

Level 2
P(V|A)

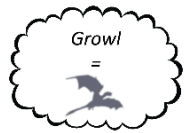

ω

Stimulus contingency
Evolution Rate

Level 1
Trialwise P(V|A)

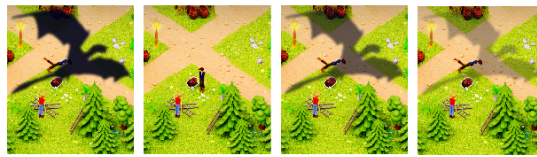

Growl   Growl   Growl   Growl

Sensory Input

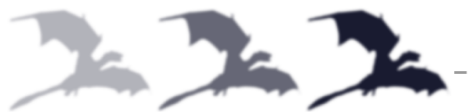

ν

Precision of Prior v.
Precision of Sensory Evidence

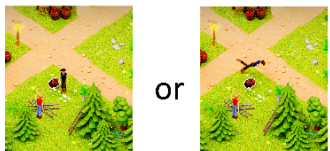 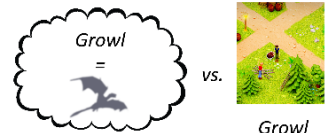

β

Inverse Decision
Temperature

Response

*Figure 1. The gameworld and the Conditioned Hallucinations task within it. a.* The character's avatar is located in the center of the screen in Panel A which represents the gameworld. The yellow box, top left in *(a)*, represents a goal to be reached. The world is available for the player to explore and they may encounter computational tasks built into the environment as they seek to find paths towards the goal. Both task performance and exploration will be analyzed using computational models. As an example of this is the implementation of the visual version of the CH task: in the lower left of *(a)*, a dragon's shadow is present. This is part of one implementation of the conditioned hallucinations task (see *(b)*). Finally, in the red box on the lower right is an owl that may serve as a target for an ongoing attention task, should it be necessary to track player attention over time. *B.* In this figure, we represent the traditional hierarchical gaussian filter (HGF) model as it pertains to the Conditioned Hallucinations (CH) game in the example gameworld. The version of the CH task demonstrated here is a visual conditioned hallucination as it is more intuitive to demonstrate in a static figure, but the auditory version of the CH task can easily be implemented as well. In the game, we would modify the traditional conditioned hallucinations task such that whenever a player hears a growl and sees a dragon shadow, they should dodge it. However, if they do not see the shadow, they should not jump. As has been done successfully in other iterations of the task, the participant learns to associate the shadow with the growl and reports seeing the shadow (by dodging) even when it is absent. In the HGF, there are three levels that form an agent's perceptual model of the world in the game. Level 1 reflects the agent's belief regarding the presence/absence of the dragon shadow on any given trial. This is reflected in their decision to dodge or not. Level 2 reflects their belief that the dragon shadow is associated with the growl. Finally, level 3 represents their belief in the volatility of the association between the growl and the shadow. Based on the participant's decision to dodge or not, we can derive three important latent parameters that could act as behavior-based biomarkers of various psychiatric disorders: decision noise ($\beta^{-1}$), learning rate of the auditory-visual stimulus associations ($\omega$), and weighting of prior expectations ($\nu$).